# Anomalous transverse optic phonons in SnTe and PbTe - revisited

Zehua Li,[1,†,*] Shasha Li,[2,3,†] John-Paul Castellan,[4] Rolf Heid,[1] Yu Xiao,[5] Li-Dong Zhao,[5] Yue Chen,[2] Frank Weber[1]

[1] *Institute for Quantum Materials and Technologies, Karlsruhe Institute of Technology, 76021 Karlsruhe, Germany*
[2] *Department of Mechanical Engineering, The University of Hong Kong, Pokfulam Road, Hong Kong SAR, China*
[3] *School of Science, Nanjing University of Posts and Telecommunications, Nanjing 210023, China*
[4] *Laboratoire Léon Brillouin (CEA-CNRS), CEA-Saclay, F-91911 Gif-sur-Yvette, France*
[5] *School of Materials Science and Engineering, Beihang University, Beijing 100191, China*

[†] these authors contributed equally to this work
[*] current address: Department of Inorganic Chemistry, Fritz-Haber-Institut der Max-Planck-Gesellschaft, 14195 Berlin, Germany

**Abstract:**

We present a study of the soft transverse optic phonon mode in SnTe in comparison to the corresponding mode in PbTe using inelastic neutron scattering and *ab-initio* lattice dynamical calculations. In contrast to previous reports our calculations predict that the soft mode in SnTe features a strongly asymmetric spectral weight distribution qualitatively similar to that found in PbTe. Experimentally, we find that the overall width in energy of the phonon peaks is comparable in our neutron scattering spectra for SnTe and PbTe. We observe the well-known double-peak-like signature of the TO mode in PbTe even down to T = 5 K questioning its proposed origin purely based on phonon-phonon scattering. The proximity to the incipient ferroelectric transition in PbTe likely plays an important role not included in current models.

**Introduction**

Thermoelectric materials are interesting for energy applications as they can transform heat into useful electricity[1,2]. The energy transformation efficiency of a thermoelectric material is determined by the figure of merit, $ZT = \sigma S^2 T/(\kappa_L + \kappa_E)$, where $\kappa_L$ and $\kappa_E$ denote the lattice and electronic thermal conductivity, respectively; $\sigma$ represents the electrical conductivity; $S$ is the Seebeck coefficient; and $T$ is the absolute temperature. An efficient thermoelectric material must have a low thermal conductivity. Thus, a detailed understanding of the lattice dynamical properties is important to design new materials with an as-small-as-possible lattice thermal conductivity for improved thermoelectric performance.

Rock-salt compounds SnTe and PbTe feature a low thermal conductivity and are among the most efficient thermoelectric materials[2-5]. While lattice dynamics were investigated early on by inelastic neutron scattering (INS)[6], more recent experiments[7-10] revealed highly anomalous phonon properties in PbTe, i.e., a double-peak-like structure of the zone center transverse optic (TO) phonon mode as well as an avoided crossing of the longitudinal acoustic (LA) and TO phonon modes. These observations were explained by various *ab-initio* lattice dynamical calculations[7,11,12] which assigned them to anharmonic phonon-phonon scattering. Another scenario proposes that a locally broken symmetry produces the TO mode's double-peak-like structure [10] similar to findings in other IV-VI compounds [13]. Generally it is assumed that the unusual lattice dynamics play an important role with regard to the very low lattice thermal conductivity of $\kappa_{lat} = 2 \text{ Wm}^{-1}\text{K}^{-1}$ in PbTe[14]. While PbTe is an incipient ferroelectric with an estimated negative transition temperature $T_c$ [15], the TO zone center phonon softens completely in isostructural SnTe [16] where the transition temperature depends sensitively on the charge carrier concentration [17,18]. Though it is widely accepted that SnTe also features anharmonic lattice dynamics, previous experimental [8] and theoretical [12] investigations did not report the presence of a double-peak-like spectral weight distribution of the TO zone center mode.

Here, we report a combined theoretical and experimental study of the TO zone center mode in SnTe and PbTe. Whereas our *ab-initio* lattice dynamical calculations and inelastic neutron scattering experiments nicely demonstrate the double-peak-like features observed previously in PbTe, we calculate a similar asymmetric spectral weight distribution for SnTe at room temperature overlaid with the strong temperature dependence of the main phonon energy reminiscent of the soft mode behavior. Experimentally, we demonstrate that the phonon scattering in PbTe and SnTe is similar if we take into account the different values of the finite (SnTe) and estimated (PbTe) ferroelectric transition temperatures. Moreover, the double-peak-like feature in PbTe does not vanish even at T = 5 K questioning a purely phonon-phonon scattering based origin.

**Calculation**

Phonon dispersions of SnTe and PbTe were calculated by extracting the harmonic interatomic force constants (IFCs) based on the finite-displacement approach using a 4×4×4 supercell. Self-consistent phonon (SCPH) theory was employed to obtain the anharmonic phonon frequency by considering the fourth order anharmonicity non-perturbatively. By applying perturbation theory to the SCPH results, frequency shifts and phonon lifetime due to the third order anharmonic terms were calculated [19]. The cubic and quartic IFCs Φ necessary for the SCPH and perturbation calculations were extracted based on the compressive sensing approach using the least absolute shrinkage and selection operator (LASSO) technique [20], which works well for both simple systems such as Si and $Mg_2Si$ [21] and complex system like $SrTiO_3$ [19]. More computation details are given in Appendix A.

**Experiment**

High-quality SnTe and PbTe crystals were synthesized using a modified vertical Bridgman method [22]. High-purity elemental constitutes were measured and loaded into carbon-coated conical silicon tubes and the crystals were grown in a temperature gradient from 1273 K to 973 K at a slow rate of 1 K/h. Finally, SnTe and PbTe crystals with diameter of ~ 12 mm and length of ~ 30 mm were obtained. They were characterized with x-ray diffraction on a Huber four-circle diffractometer able to achieve temperatures down to T = 5 K. INS experiments were performed on the 1T triple-axis spectrometer at the ORPHEE reactor at Laboratoire Léon Brillouin (LLB, CEA Saclay) using doubly focusing PG002 monochromator and analyzer. The final energy was fixed at 8 meV. The single crystals PbTe and SnTe weighing 5 g and 3 g, respectively, were mounted in a closed-cycle refrigerator allowing measurements down to T = 5 K. The experimental resolution was 0.9 meV in the neutron scattering experiments for elastic scattering (FWHM). All measurements were carried out in the 110-001 scattering plane for temperatures 5 K ≤ T ≤ 300 K. The wave vectors are given in reciprocal lattice units (r.l.u.) of ($2\pi/a$, $2\pi/b$, $2\pi/c$), with lattice constants $a = b = c$ = 6.32 and 6.45 Å for SnTe and PbTe, respectively.

**Results**

Figure 1 shows the calculated phonon spectra of the zone center TO mode in SnTe and PbTe at different temperatures. At low temperature, T = 75 K, we find well-defined peaks [Fig. 1(a)], which broaden out towards higher energies on approaching room temperature [Fig. 1(b)-(d)]. At high temperatures, the spectra are dominated again by a sharp feature [Fig. 1(e)] which emerges gradually on the high-energy side of the broad spectral weight distribution calculated for room temperature. More specifically, the spectra for PbTe display the well-known double-peak-like feature at intermediate temperatures [7,8] and agree well with previous calculations by one of us [11]. Results for SnTe emphasize the TO soft mode character with peak energies ranging from 7.7 meV at high [Fig. 1(e)] to 0.8 meV at low temperatures [Fig. 1(a)]. We note that the peak energy becomes imaginary for smaller temperatures in agreement with the observed structural-ferroelectric phase transition in SnTe. However, our calculations reveal also strongly asymmetric spectral weight distribution at intermediate temperatures [Figs. 1(b)-(d)]. In particular, the width of the spectral weight distribution calculated for room temperature is similar for SnTe and PbTe.

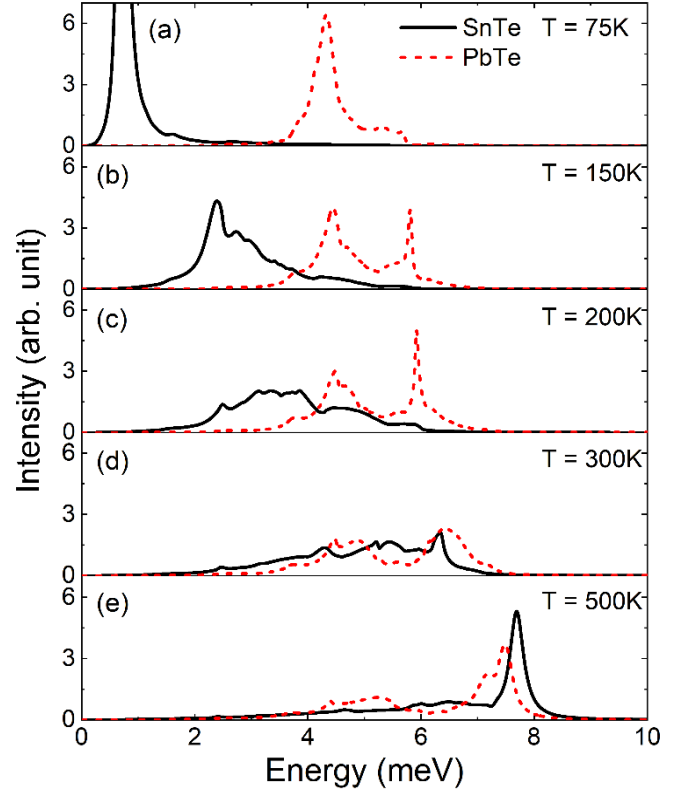

**FIG. 1** (a)-(e) Calculated spectrum of the TO mode at zone center $Q = (1, 1, 3)$ in SnTe (black solid lines) and PbTe (red dashed lines) at selected temperatures 75 K ≤ T ≤ 500 K.

The theoretical predictions motivated us to perform phonon measurements in SnTe and PbTe using thermal neutron triple-axis spectroscopy. We employed a low final energy of $E_f$ = 8 meV indispensable to separate low-energy phonon intensities from strong Bragg scattering at the zone center. Raw data taken at $Q = (1, 1, 3)$ feature strong elastic scattering and low-energy inelastic scattering on top of an experimental background [inset of Fig. 2(a)]. Approximating a Gaussian to the elastic scattering demonstrates that scattering intensities at $E \geq$ 1.75 meV are not contaminated by elastic scattering. Hence, we consider only scattering contributions at E ≥ 1.75 meV from which we subtract a constant experimental background [dashed line in inset of Fig. 2(a)]. A corresponding procedure was used to subtract the background for the spectra in PbTe.

The resulting background-subtracted data were corrected for the Bose factor. The data reveal an intense low-energy phonon, the TO mode, followed by weak peaks at 11 meV and 17 meV [Fig. 2(a)] in SnTe. We assign the latter to the LO mode and argue that the 11 meV originates from a small polycrystalline part of our sample reflecting phonon density of states (PDOS) properties (see Appendix B and Fig. 8 for more details). In the data shown in the following graphs, we subtracted the PDOS-related peak to simplify the analysis.

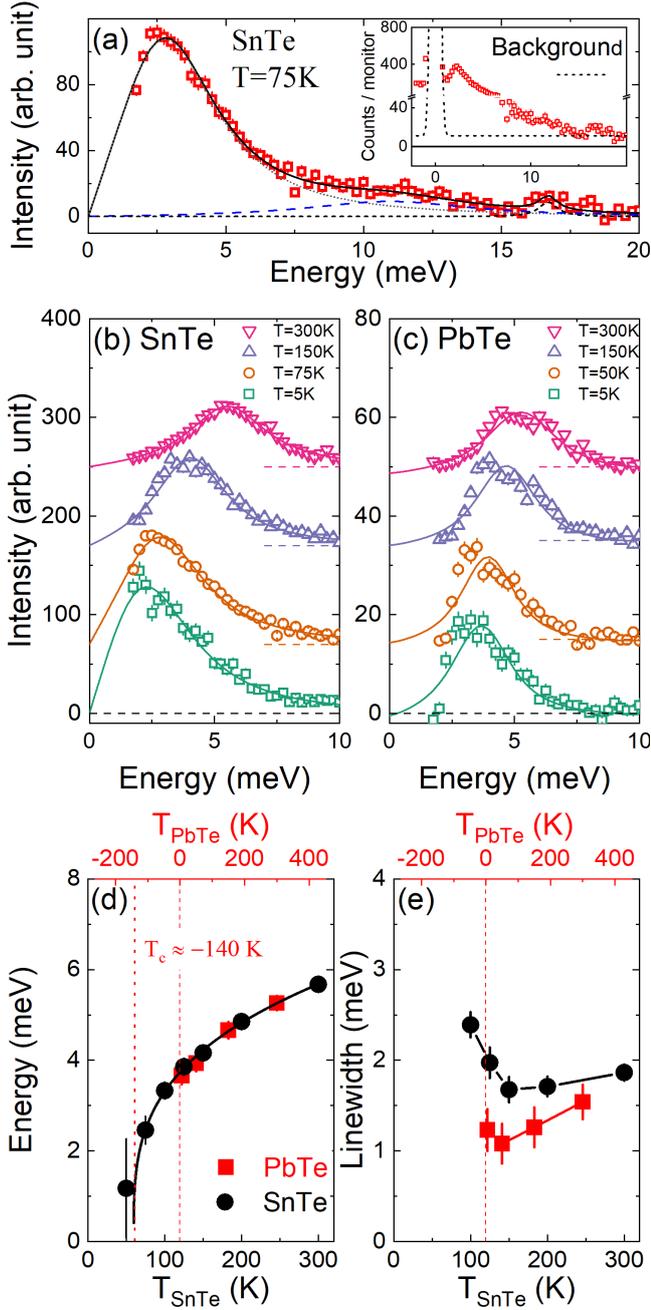

FIG. 2 INS data at $Q = (1, 1, 3)$ in SnTe and PbTe. (a) Background subtracted and Bose corrected INS data for SnTe at T = 75 K (squares). Solid line is a fit to the data consisting of a DHO function for the inelastic peaks at 3 meV (dotted line) and 17 meV (dashed line) convoluted with the resolution function along with a Gaussian (blue dashed line) representing a powder-like scattering contribution (see text). The inset in (a) shows the corresponding INS raw data (squares) along with the estimated experimental background (dashed line). (b)(c) Background subtracted and Bose corrected data taken at $Q = (1,1,3)$ for (b) SnTe and (c) PbTe at selected temperatures. Solid lines are fits to the data as discussed above. Data are offset vertically for clarity. (d) Observed phonon energies and (e) linewidths as function of temperature for SnTe (black dots, bottom scale) and PbTe (red squares, top scale). The line in (d) is a power-law fit to the SnTe data with T ≥ 75 K of the form $E_{phon} \sim (T-T_0)^\delta$ yielding $\delta = 0.3 \pm 0.01$ and $T_0 = (59.6 \pm 3.4)$ K.

We approximated the TO phonon peak with a DHO function convoluted with the Gaussian experimental resolution for both SnTe [Fig. 2(b)] and PbTe [Fig. 2(c)]. Within our experimental accuracy, the TO mode softens completely in SnTe [black dots in Fig. 2(d)] in agreement with a ferroelectric transition temperature $T_c \approx 60$ K deduced from x-ray diffraction (see Appendix C and Fig. 9). Meanwhile, the approximated phonon energy for PbTe decreases from 5.3 meV at room temperature to 3.7 meV at T = 5 K [red squares in Fig. 2(d)]. In Figure 2(d) we show the temperatures for PbTe in a way that the temperature dependence of the observed phonon energies overlaps with that found in SnTe. Thereby, we estimate a critical temperature of the incipient ferroelectric transition in PbTe of $T_c \approx -140$ K [vertical dashed line in Fig. 2(d)], which is consistent with previous reports ($T_c = -151$ K [23] ; $T_c = -135$ K [24]). Using the same arrangement for the temperature axes to plot the obtained phonon linewidths, we find that the values for the TO mode in PbTe are about 25% smaller than those for SnTe in the comparable temperature range, *i.e.*, $T_{SnTe} \geq 125$ K [Fig. 2(e)]. The TO linewidth in SnTe increases on further cooling towards $T_c$.

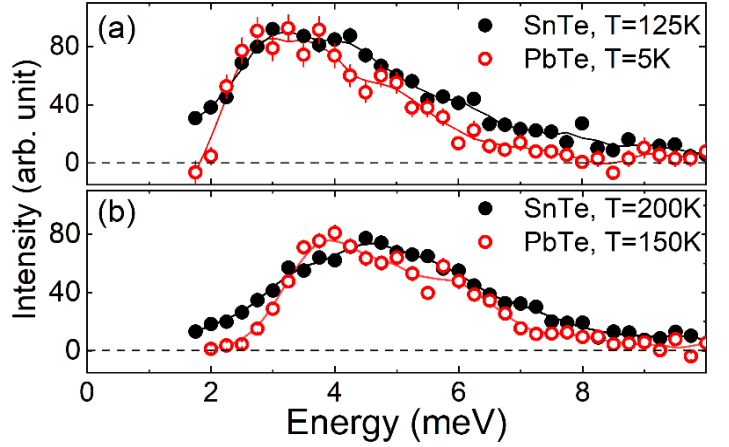

**Figure 3.** (a)(b) Comparison of phonon spectra for SnTe and PbTe taken at equivalent temperatures based on the temperature scaling relationship from Fig. 2(d) and discussed in the text. Lines are guides to the eye. Count rates for PbTe were up-scaled by a factor of 4.88 for easy comparison.

The presented peak analysis shows that the properties of the TO mode are quite similar when we take into account that PbTe is much farther away from the corresponding structural phase transition compared to SnTe. However, the analysis using a DHO peak function also reveals clear short-comings, in particular in the case of PbTe [Fig. 2(c)]: the overall shape of the spectral weight distribution is not well-described. The data feature a much steeper increase at the low-energy than at the high-energy side of the peak and suggest a double-peak-like spectral weight distribution in agreement with calculations (see Fig. 1) and previous work [7,25]. Comparing data taken at corresponding temperatures shows that the spectral weight distribution is more structured in the case of PbTe (Fig. 3). However, the range of energies over which the phonon intensities are distributed are about the same in both compounds. This is consistent with our calculations where we see generally more structured phonon spectra in PbTe [Fig. 1(b)(c)(d)], but similarly wide intensity distribution.

We find reasonable agreement between the calculated and observed spectra at room temperature for both compounds [Figs. 4(a)(d)]. Yet, this changes at lower temperatures. In SnTe, the calculated spectra soften much stronger than it is observed. Furthermore, theory predicts a drastic narrowing while the measured TO mode even broadens at $T \leq 100$ K [Figs. 4(b)(c) and 2(e)]. Calculations close to the ferroelectric phase transition are expected to be very temperature sensitive. Whereas the estimated transition temperature [see Appendix A, Fig 5] is in reasonable agreement, the calculated phonon softening follows a mean-field-like power law, *i.e.*, $E_{TO} \sim (T-T_c)^\delta$ with $\delta \approx 0.5$ (Fig. 5). Experimentally, we observe $\delta \approx 0.3$ [data in Fig. 2(d)]. Thus, the phonon energy at intermediate temperatures, e.g., at T = 150 K [Fig. 4(b)], and close to the phase transition [Fig. 4(c)] is underestimated. The softening in PbTe is more moderate and the agreement with theory is reasonable. However, the predicted narrowing at low temperatures is not observed as well [Figs. 4(e)(f) and 2(e)].

The narrowing at low temperatures is essentially built-in to our model as well as to other anharmonic ones previously used [7,8]. Such models are based on low-temperature harmonic calculations, i.e., having by default a zero intrinsic phonon line width at zero Kelvin. The calculated broad spectral weight distribution at elevated temperatures purely originates from phonon-phonon scattering which is frozen out at low temperatures. However, a structural phase transition itself can be regarded as an anharmonic double-well potential yielding finite phonon lifetimes. This possible effect of the phase transition on the phonon line width is typically not taken into account in current models. From the comparison with SnTe [Fig. 2(e)], one can speculate that the linewidth of the TO mode in PbTe would even increase if one could get closer to the ferroelectric/structural phase transition. Certainly, the (incipient) ferroelectric phase transition plays an important role for the lattice dynamics in both materials. In particular, the observed TO mode in PbTe at base temperature of T = 5 K questions a purely phonon-phonon-scattering based origin of the broad and asymmetric spectral weight distribution.

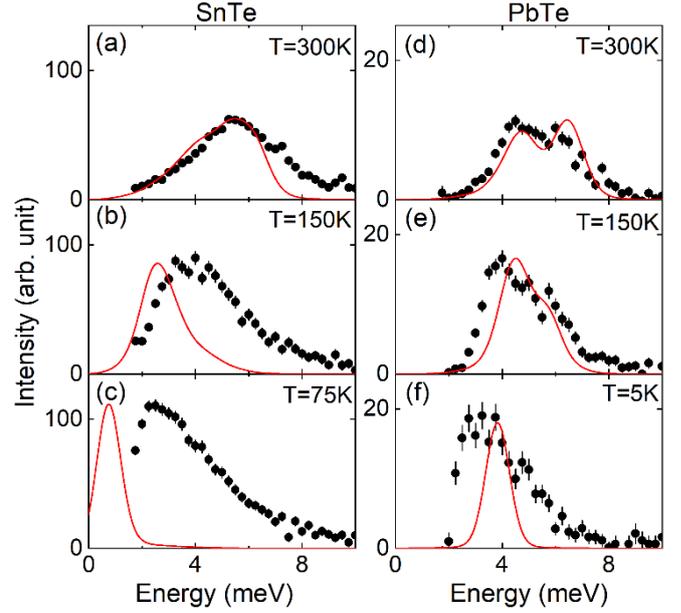

**Figure 4.** Comparison of calculated (lines) and measured phonon spectra (dots) for various temperatures in (a)-(c) SnTe and (d)-(f) PbTe. Calculated spectra have been convoluted with the experimental resolution and scaled to match the respective peak intensities.

**Discussion**

The current interest in the lattice dynamics of IV-VI semiconductors is based on the good thermoelectric performance of PbTe [7,26,27] and related materials [5,22,28]. The TO zone center mode attracts particular attention as anharmonic behavior was recognized early on [6]. Our experimental observations are in-line with several INS reports [7,8,10] on the double-peak-like spectral weight distribution of the TO mode in PbTe. For SnTe, we can confirm the previously estimated [16] full softening of the TO mode at $q$ = 0. For both compounds, pair distribution function investigations (PbTe [26], SnTe [29]) reported local distortions related to an off centering of the Pb (Sn) atom inconsistent with the globally cubic crystalline structure at high temperatures. In a subsequent INS study of PbTe [10], the double-peak-like spectral weight distribution at the zone center was interpreted as the emergence of a new mode on heating because of these local distortions. On the other hand, INS investigations along with *ab-initio* molecular dynamics (AIMD) simulations [7,8] could explain the TO mode's satellite peak due to phonon dispersion nesting between the LA and TO modes over a large phase space. In more detail, the energy of the satellite corresponds to $\Delta E = E_{TO} - E_{LA}$ [8], which is very similar to our own calculations [see Appendix A and Fig. 6(b)]. This anomalous phonon-phonon interaction between longitudinal and transverse modes and the concurrent strong damping of the heat-carrying LA mode was held responsible for the low lattice thermal conductivity [7]. The absence of a similar satellite peak in SnTe was assigned to a less pronounced phonon nesting effect [8]. Following work [9] rationalized an asymmetry in the 1st nearest neighbor peak of the radial distribution function due to large-amplitude anharmonic phonons for both PbTe and SnTe without the need to invoke local distortions.

What sets our results in PbTe apart is that we observe the double-peak-like phonon structure – or satellite peak – even at a low temperature of T = 5 K [Fig. 4(f)]. The models presented for both, the atomic off centering [26,29] and the strong anharmonic LA-TO interaction [7-9], explain the anomalous effects only on heating. Following [10,26,29], the degeneracy of the TO mode in the cubic environment should be partially removed but only for $T > 100$ K (PbTe) and $T > 300$ K (SnTe). In contrast, our results suggest that the presence (SnTe) or proximity (PbTe) of the ferroelectric phase transition must be taken into account to explain the low temperature lattice dynamics.

The nature of the ferroelectric phase transition in SnTe but also in closely related GeTe has been the subject of numerous investigations. For the latter compound, it was argued [13] that local distortions of the Ge-Te bond essentially remain unchanged on heating through the ferroelectric(rhombohedral)-to-paraelectric(cubic) phase transition. This could explain a splitting of the TO zone center mode even in the paraelectric cubic phase which was reported via Raman scattering [13]. However, this proposal was rejected in other publications suggesting a displacive transition in GeTe [30,31]. For SnTe, recent reports agree on a displacive-type phase transition evidenced by phonon softening above and subsequent hardening below $T_c$ [16]. The results in [16] were extrapolated to $q$ = 0 because inelastic x-ray scattering cannot be used at momenta with strong elastic scattering, *e.g.*, Bragg scattering. Hence, our results in SnTe represent a more accurate and direct determination of the phonon softening in SnTe revealing a non-mean-field temperature dependence of the phonon energy proportional to $(T-T_s)^{0.3}$ [Fig. 2(d)].

We have discussed above that our anharmonic calculations as well as those of others are based on a stable harmonic ground state calculations and, thereby, neglect possible effects arising from the presence of an anharmonic energy potential at low temperature. Recently, the phase transition in SnTe and its lattice dynamics were investigated by *ab-initio* calculations in the *stochastic self-consistent harmonic approximation* (SSCHA) [12]. This method takes into account anharmonic effects due to phonon-phonon scattering as function of temperature as well as quantum effects important at low temperatures. The calculations are able to reproduce the salient phonon features in PbTe at room temperature along with a reasonable transition temperature in SnTe. Furthermore, strong anharmonic broadening of the TO mode in SnTe is reported and the displacive nature of the phase transition supported. In particular, the anharmonic double-well potential underlying the ferroelectric transition is scrutinized, However, there are no details about the temperature dependence of the TO mode's linewidth. But it seems that this model is the most likely one to be able to reproduce both the high and low temperature lattice dynamics in PbTe and SnTe.

**Conclusion**

In summary, we studied the evolution of the zone center TO mode in PbTe and SnTe by computational methods and high energy-resolution neutron spectroscopy. We demonstrate that the TO spectral weight at room temperature is as broadly distributed in SnTe as in PbTe evidenced by the linewidths obtained from a single-peak analysis. Experiments in PbTe reveal a double-peak-like spectral weight distribution even at T = 5 K and are in contrast to various proposed models. Rather, the comparison with SnTe suggests that the mechanism of the ferroelectric phase transition plays a decisive role for the low temperature lattice dynamical properties. We discuss that a successful model description needs to address anharmonic and quantum effects on equal footing.

**Acknowledgements**

Z.L. was supported by the Helmholtz-OCPC Postdoc Program. S.L. and Y.C. are grateful for the financial support of Research Grants Council of Hong Kong (G-HKU704/17) and the research computing facilities offered by ITS, HKU. Z.L. and F.W. were supported by DAAD-PPP project 57391742. S.L. acknowledges the National Natural Science Foundation of China (Grant No. 12104234), Natural Science Foundation of Jiangsu Province (Grant Nos.BK20210578 and 20KJB140004) and NUPTSF (Grant No. NY220096).

**Appendix A - Calculation details**

Phonon dispersions of SnTe and PbTe were calculated by extracting the harmonic interatomic force constants (IFCs) based on the finite-displacement approach using a 4×4×4 supercell. Self-consistent phonon (SCPH) theory was employed to obtain the anharmonic phonon frequency by considering the fourth order anharmonicity non-perturbatively. By applying perturbation theory to the SCPH results, frequency shifts and phonon lifetime due to the third order anharmonic terms were calculated [19]. The cubic and quartic IFCs Φ necessary for the SCPH and perturbation calculations were extracted based on the compressive sensing approach using the least absolute shrinkage and selection operator (LASSO) technique [20], which works well for both simple systems such as Si and $Mg_2Si$ [21] and system $SrTiO_3$ [19]. To solve the LASSO equation:

$$\widetilde{\Phi} = arg \min_{\Phi} \|A\Phi - F\|_2^2 + \lambda\|\Phi\|_1 \qquad (1)$$

ab initio molecular dynamics (AIMD) simulations in the canonical ensemble (NVT) for 4×4×4 supercells of SnTe and PbTe at 300 K were conducted for 2000 steps with a time step of 2 fs. A 1×1×1 Monkhorst-Pack grid [32] was used during AIMD simulations. From trajectories of the last 1000 steps, we sampled 40 atomic configurations equally spaced in time. For each configuration, all the atoms were displaced by 0.1 Å in random directions. The atomic forces for these configurations were calculated based on density functional theory (DFT), from which the displacement matrix $A$ and the force vector $F$ were constructed. The LASSO equation was solved using the split Bregman algorithm, and the optical value of λ was selected from the four-fold cross-validation score.

When calculating the force vector $F$, ab initio DFT calculations were performed using the Vienna ab initio simulation package (VASP) [33] and the energy cutoff was set to 400 eV. The electron-ion interactions were treated with the projector augmented wave (PAW) method [34] and the exchange-correlation interactions were treated by the Perdew-Burke-Ernzerhof (PBE) functional [35]. A 3×3×3 Monkhorst-Pack grid [32] was used for the 4×4×4 supercell. The splitting between the longitudinal and transverse optical modes (LO–TO splitting) at the Γ point was considered by calculating the Born effective charges and dielectric constants using density functional perturbation theory (DFPT).

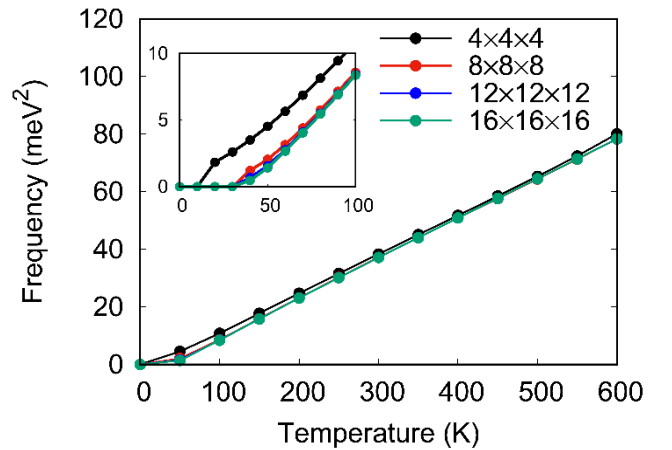

**FIG. 5** Temperature dependent phonon frequency ($\Omega_q^2$) of the TO mode at Γ point of rock salt SnTe obtained using different $q_1$-point densities in SCPH calculations.

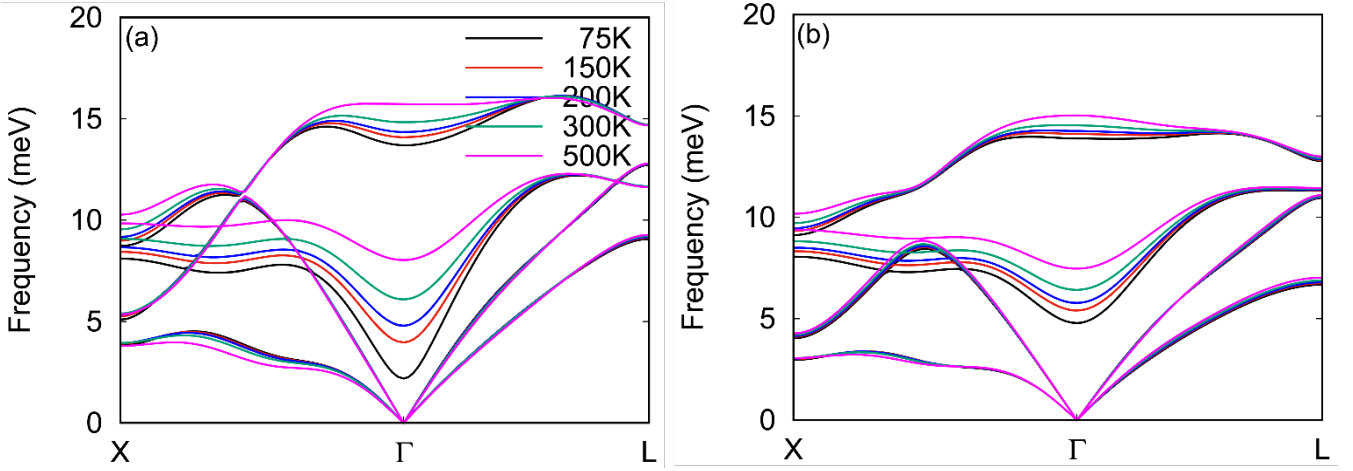

**FIG. 6** Theoretical phonon dispersions of rock salt (a) SnTe and (b) PbTe calculated from SCPH.

With the quartic IFCs $\Phi$ calculated above, anharmonic phonon frequencies $\Omega_q$ can be obtained self-consistently. Assuming that the polarization vectors and internal coordinates are not altered by anharmonic effects, self-consistent phonon equation can be expressed as [36]:

$$\Omega_q^2 = \omega_q^2 + 2\Omega_q I_q \qquad (2)$$

$$I_q = \sum_{q_1} \frac{\hbar \Phi(q;-q;q_1;-q_1)}{4\Omega_q \Omega_{q_1}} \frac{[2n(\Omega_{q_1})+1]}{2} \qquad (3)$$

where $n(\omega) = (e^{\beta \hbar \omega} - 1)^{-1}$ is the Bose-Einstein distribution function. Based on the convergence test [Fig. 5], a 12×12×12 $q_1$-point mesh was used in the SCPH calculation for rock salt SnTe and PbTe.

To include the cubic anharmonicity, we calculated the bubble self-energy $\Sigma_q^{bubble}(\Omega_q)$. The frequency shift due to cubic anharmonicity was estimated by $\Delta_q^{bubble} = -\frac{1}{\hbar} Re \sum_q^{bubble}(\Omega_q)$, and the phonon lifetime was estimated as $\tau_q = 1/(2\Gamma_q^{bubble}(\Omega_q))$, where $\Gamma_q^{bubble} = \frac{1}{\hbar} Im \sum_q^{bubble}(\Omega_q)$.

Phonon spectra $\chi_q(\omega)$, which has a direct connection with INS experiments, were calculated:

$$\chi_q(\omega) \propto \frac{2\Omega_q^2 \Gamma_q^{bubble}(\omega)}{[\omega^2 - \Omega_q^2 - 2\Omega_q \Delta_q^{bubble}(\omega)]^2 + 4\Omega_q^2 \Gamma_q^{bubble}(\omega)^2} \qquad (4)$$

For convergence test, $q_1$-point meshes of 4×4×4, 8×8×8, 12×12×12, and 16×16×16 were used in SCPH to calculate the phonon frequency of the soft mode of cubic SnTe at Γ point. Figure 5 shows that a 12×12×12 $q_1$-point mesh is needed to obtain convergence, and the $T_c$ of R3m-to-cubic phase transition is about 30 K. To be consistent, the same $q_1$-point density is used in SCPH calculation for PbTe.

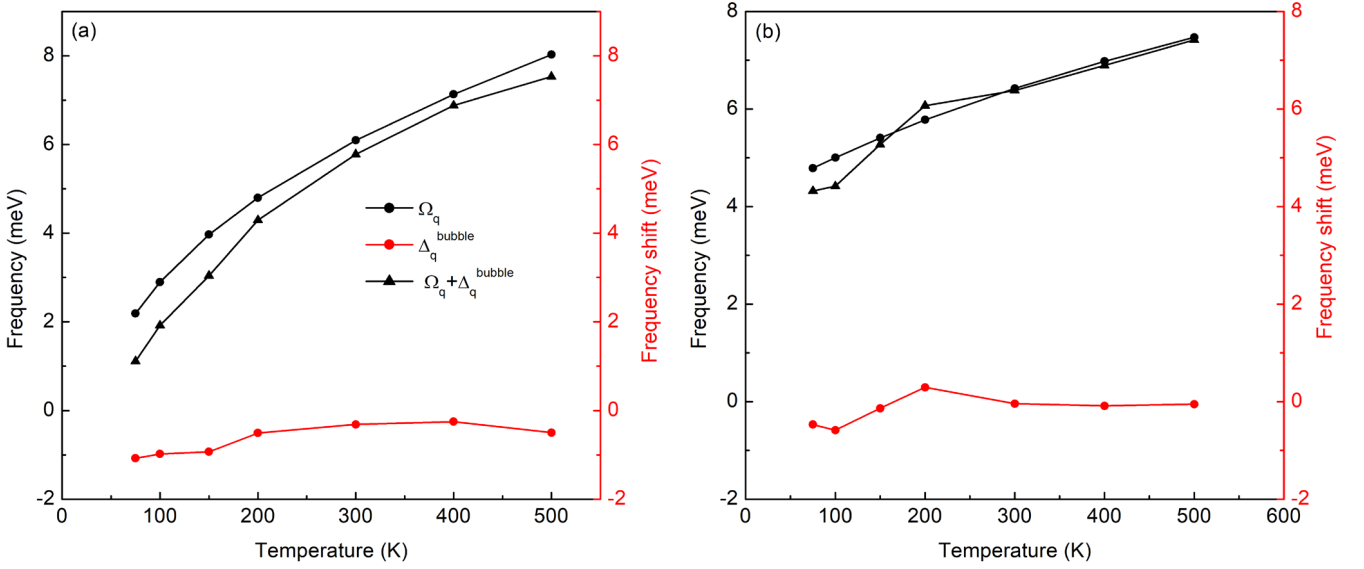

**FIG. 7** Phonon frequency of the TO mode at Γ point ($\Omega_q$, black dots) for rock salt (a) SnTe and (b) PbTe calculated from SCPH theory considering only the quartic terms, frequency shifts from the bubble self-energy considering the cubic anharmonicity ($\Delta_q^{bubble}$, red dots), and the overall phonon frequency ($\Omega_q+\Delta_q^{bubble}$, blue dots).

The anharmonic phonon dispersions of SnTe and PbTe obtained from the SCPH calculations are given in Figure 6. It is seen that the optical modes stiffen due to quartic anharmonicity, especially those around the zone center. The avoided LA-TO crossing of PbTe at elevated temperatures can be also observed, which is consistent with previous report [37].

The results of $\Omega_q + \Delta_q^{bubble}$ for the TO mode at Γ point are shown in Figure 7. It is seen the $\Delta_q^{bubble}$ term is negative and almost negligible.

**Appendix B - Calculated phonon density of states**

In order to investigate the origin of the small additional peak around 11 meV in our data for SnTe, we compare experimental and calculated phonon density of states (PDOS) in Figure 8. The neutron-weighted PDOS (red line in Fig. 8) was calculated as $\sum_k \frac{\sigma_k}{m_k} \cdot g_k$, where $g_k$ is the calculated partial PDOS of element $k$ and $\sigma_k$ and $m_k$ are the corresponding neutron scattering cross section and atomic mass. This result is compared to the expected $\boldsymbol{Q}$-averaged intensities for a narrow range of absolute $\boldsymbol{Q}$ values close to that of $\boldsymbol{Q}$ = (1,1,3), i.e., $|\boldsymbol{Q}_{113}|$ = 3.297 Å$^{-1}$. To this end, we performed phonon structure factor calculations on a regular three dimensional grid with spacing of $0.05 \times \frac{2\pi}{a}$, $0.05 \times \frac{2\pi}{b}$ and $0.05 \times \frac{2\pi}{c}$ along the three axes of the reciprocal unit cell. Individual phonons for a particular wave vector $\boldsymbol{Q}$ were simulated by resolution-limited peaks using the calculated structure factors to scale the peak amplitudes. Subsequently, phonon intensities for wavevectors in a very small $|Q|$ range were averaged. Results for the range $3.296 \leq |Q| \leq 3.298$ Å$^{-1}$ ($|Q_{113}| = 3.297$ Å$^{-1}$) are shown as blue solid line in Figure 8.

The calculated PDOS shows reasonable agreement with the experimental PDOS from INS [9] [open squares in Fig. 8]. Moreover, the calculations show that the $Q$-averaged intensity for $|Q_{113}| \approx 3.297$ Å$^{-1}$ closely resembles the overall PDOS. In analogy to our data analysis shown in Figure 2(a), we approximate a DHO function to the PDOS and obtain a peak energy of approximately 11 meV very close to the value we derive for the additional peak in our INS spectra obtained at $Q = (1,1,3)$. Therefore, we assign this experimental signature to a PDOS-like signal from a small polycrystalline part in our sample. We used the deduced width of the DHO function to analyze our phonon spectra [see Fig. 2(a)]. We found that the peak is always close to 11 meV and follows the Bose statistics, which corroborates its PDOS-like origin.

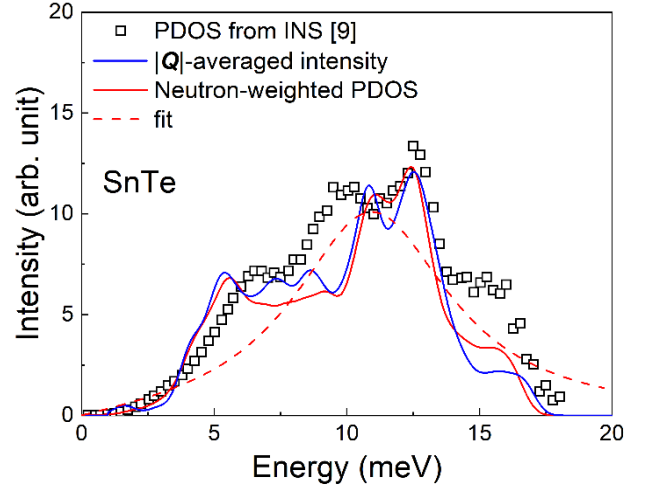

**FIG. 8** Comparison of the neutron-weighted PDOS (red solid line) with experimental results (open squares) presented by Li et al. [9]. The blue solid line denotes calculated phonon intensities which were $Q$-averaged in the range $3.296$ Å$^{-1} \leq |Q| \leq 3.298$ Å$^{-1}$ ($|Q_{(113)}| = 3.297$ Å$^{-1}$). The red dashed line represents a DHO fit to the neutron-weighted PDOS.

**Appendix C - Phase transition of SnTe**

We deduced the structural transition temperature of SnTe from single crystal x-ray diffraction. In particular, we investigated the $Q = (3,3,1)$ Bragg peak, which shows a clear increase of its intensity on cooling below temperatures of 60 K (Fig. 9). A comparable data set for PbTe demonstrates the absence of a similar transition for temperatures down to the base temperature of our setup, i.e., $T = 5$ K. Thus, we define the transition temperature of our SnTe sample to be $T_c \approx 60$ K, which is in reasonable agreement with with previous reports, i.e., $T_c$ = 42 K [38] and $T_c$ = 80 K [29]. While PbTe has no phase transition in the range 5 K ≤ $T$ ≤ 300 K, we estimate $T_c$ ≈ −140 K based on a comparison of the phonon energies in PbTe and SnTe [Fig. 2(d)], consistent with values given in previous reports, *i.e.*, −151 K [23] and −135 K [24].

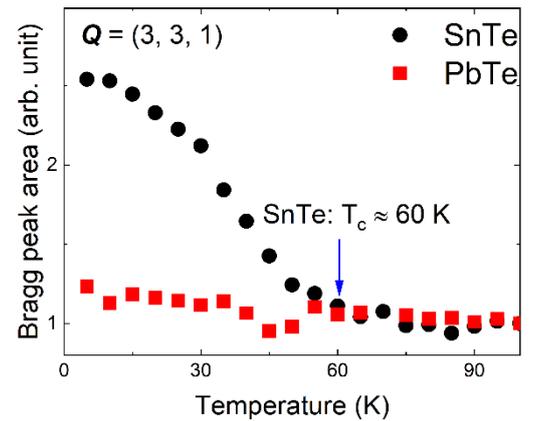

**FIG. 9** XRD intensity of the Bragg peak at $Q = (3,3,1)$ for SnTe (dots) and PbTe (squares). Results are normalized to 1 at $T = 100$ K.

**Figure captions**

**FIG. 1** (a)-(e) Calculated spectrum of the TO mode at zone center $Q = (1, 1, 3)$ in SnTe (black solid lines) and PbTe (red dashed lines) at selected temperatures 75 K ≤ T ≤ 500 K.

**FIG. 2** INS data at $Q = (1, 1, 3)$ in SnTe and PbTe. (a) Background subtracted and Bose corrected INS data for SnTe at T = 75 K (squares). Solid line is a fit to the data consisting of a DHO function for the inelastic peaks at 3 meV (dotted line) and 17 meV (dashed line) convoluted with the resolution function along with a Gaussian (blue dashed line) representing a powder-like scattering contribution (see text). The inset in (a) shows the corresponding INS raw data (squares) along with the estimated experimental background (dashed line). (b)(c) Background subtracted and Bose corrected data taken at $Q = (1,1,3)$ for (b) SnTe and (c) PbTe at selected temperatures. Solid lines are fits to the data as discussed above. Data are offset vertically for clarity. (d) Observed phonon energies and (e) linewidths as function of temperature for SnTe (black dots, bottom scale) and PbTe (red squares, top scale). The line in (d) is a power-law fit to the SnTe data with T ≥ 75 K of the form $E_{phon} \sim (T-T_0)^\delta$ yielding $\delta = 0.3 \pm 0.01$ and $T_0 = (59.6 \pm 3.4)$ K.

**FIG. 3.** (a)(b) Comparison of phonon spectra for SnTe and PbTe taken at equivalent temperatures based on the temperature scaling relationship from Fig. 2(d) and discussed in the text. Lines are guides to the eye. Count rates for PbTe were up-scaled by a factor of 4.88 for easy comparison.

**FIG. 4.** Comparison of calculated (lines) and measured phonon spectra (dots) for various temperatures in (a)-(c) SnTe and (d)-(f) PbTe. Calculated spectra have been convoluted with the experimental resolution and scaled to match the respective peak intensities.

**FIG. 5** Temperature dependent phonon frequency ($\Omega_q^2$) of the TO mode at Γ point of rock salt SnTe obtained using different $q_1$-point densities in SCPH calculations.

**FIG. 6** Theoretical phonon dispersions of rock salt (a) SnTe and (b) PbTe calculated from SCPH.

**FIG. 7** Phonon frequency of the TO mode at Γ point ($\Omega_q$, black dots) for rock salt (a) SnTe and (b) PbTe calculated from SCPH theory considering only the quartic terms, frequency shifts from the bubble self-energy considering the cubic anharmonicity ($\Delta_q^{bubble}$, red dots), and the overall phonon frequency ($\Omega_q + \Delta_q^{bubble}$, blue dots).

**FIG. 8** Comparison of the neutron-weighted PDOS (red solid line) with experimental results (open squares) presented by Li et al. [9]. The blue solid line denotes calculated phonon intensities which were $Q$-averaged in the range $3.296 \text{ Å}^{-1} \leq |Q| \leq 3.298 \text{ Å}^{-1}$ ($|Q_{(113)}| = 3.297 \text{ Å}^{-1}$). The red dashed line represents a DHO fit to the neutron-weighted PDOS.

**FIG. 9** XRD intensity of the Bragg peak at $Q = (3,3,1)$ for SnTe (dots) and PbTe (squares). Results are normalized to 1 at $T = 100$ K.